\documentclass{nature}
\usepackage{graphicx}
\makeatletter
\let\saved@includegraphics\includegraphics
\AtBeginDocument{\let\includegraphics\saved@includegraphics}
\renewenvironment*{figure}{\@float{figure}}{\end@float}
\makeatother

\usepackage{color}
\usepackage{graphicx}
\usepackage{epstopdf}
\usepackage{bm}
\usepackage{dcolumn}
\usepackage{verbatim}
\usepackage{mathbbol}
\usepackage{dcolumn}
\usepackage{amsmath}
\usepackage{placeins}
\usepackage{mathrsfs}
\usepackage{gensymb}
\usepackage{physics}
\usepackage{upgreek}
\usepackage{subcaption}
\usepackage{MnSymbol}
\usepackage{float}
\usepackage{mathrsfs}

\let\Im\undefined

\DeclareMathOperator{\Im}{\mathscr{I}}

\usepackage{lineno}

\usepackage[]{multibib}
\newcites{M}{References}

\usepackage[normalem]{ulem}

\title{Lightwave  topology for strong-field valleytronics}

\author{\'A. Jim\'enez-Gal\'an$^{1}$, R. E. F. Silva$^{1,2}$, O. Smirnova$^{1,3}$, \& M. Ivanov$^{1,4,5}$}

\begin{document}

\maketitle

\begin{affiliations}
 \item Max-Born-Institute, Berlin, Germany.
 \item Department of Theoretical Condensed Matter Physics, Universidad Aut\'onoma de Madrid, Spain
 \item Technische Universit\"at Berlin, Berlin, Germany.
 \item Department of Physics, Humboldt University, Berlin, Germany.
 \item Blackett Laboratory, Imperial College London, London, United Kingdom.
\end{affiliations}

\begin{abstract}

\end{abstract}

\textbf{
Modern light generation technology offers extraordinary capabilities for sculpting light pulses, 
with full control 
over individual electric field oscillations within each laser cycle~\cite{Wirth2011, Kfir2014, Eckle2008}. 
These capabilities are at the core of lightwave electronics~\cite{Goulielmakis2007,Krausz2009, Krausz2014, Hohenleutner2015, Wolter2015}-- the dream of ultrafast lightwave control over electron dynamics in solids, on a few-cycle to 
sub-cycle timescale, aiming at information processing 
at tera-Hertz to peta-Hertz rates.  
Here we show a robust and general approach to ultrafast, valley-selective electron excitations 
in two-dimensional materials~\cite{Schaibley2016}, by controlling the sub-cycle structure of non-resonant 
driving fields at a few-femtosecond timescale. 
Bringing the frequency-domain concept of topological Floquet systems~\cite{Oka2009,Lindner2011} to the few-femtosecond time domain, 
we develop a transparent control mechanism in real space and 
an all-optical,  non-element-specific method to coherently write, manipulate and 
read selective valley excitations using fields carried in a wide range of 
frequencies, on timescales orders of magnitude shorter than valley lifetime,
crucial for implementation of valleytronic devices~\cite{Vitale:2018aa}.
}

\begin{figure}
\centering
\includegraphics[width=\linewidth]{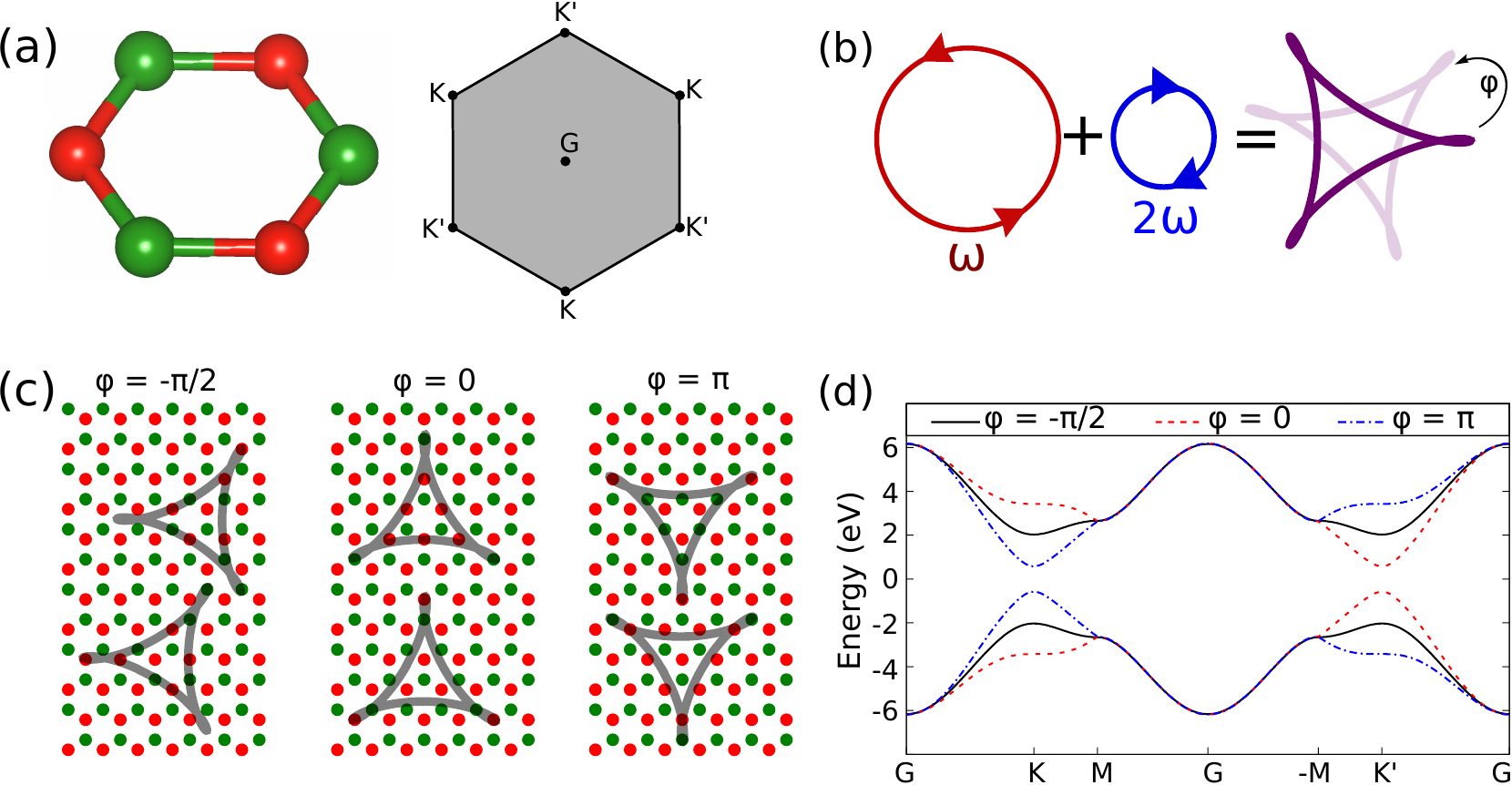}
\caption{\textbf{Light-induced modification of the band structure with tailored field}.  (a) 2D hexagonal lattice with broken inversion symmetry in real space (left; red and green represent two different atoms) and reciprocal space, with valleys $\mathbf{K}$ and $\mathbf{K'}$ (right). (b) The trefoil Lissajous figure 
generated by the field has the symmetry of the sub-lattice and 
can be rotated by changing the two-color phase $\varphi$. (c) Depending on the field orientation (grey trefoil), 
the two atomic sites are addressed differently. 
For $\varphi=-\pi/2$, the field interacts with both atoms 
equally and the bands show 
valley-degeneracy (d, black solid line). For $\varphi=0$, the field interacts with the two types of atoms differently
(note how the two atoms inside the trefoil are now not interchangeable, irrespective of where the field is placed 
in the lattice). This lifts the valley degeneracy (d, red dashed line). For $\varphi=\pi$, the situation is reversed (d, blue dashed dotted line).
\label{fig:bandmodification}}
\end{figure}

Two-dimensional graphene-like systems with broken inversion symmetry, 
such as monolayer hexagonal boron nitride (hBN) 
or transition metal dichalcogenides (TMDs), are candidates for next generation quantum 
materials due to their high carrier mobility and, especially, to their valley 
degree of freedom~\cite{Schaibley2016}, 
with potential applications in quantum information processing. 
Valleys are local minima in the crystal band structure corresponding to different crystal momenta; 
in 2D hexagonal lattices they are located at the 
$\mathbf{K}$ and $\mathbf{K}' = -\mathbf{K}$ points of the Brillouin zone (Fig.~\ref{fig:bandmodification}a). 
Selective excitation of $\mathbf{K}$ or $\mathbf{K}'$ can be achieved  using weak circularly polarized field resonant 
with the direct band gap of the material~\cite{Mak:2012aa}: it couples to either $\mathbf{K}$ or $\mathbf{K'}$ depending on 
light's helicity (the optical valley selection rule~\cite{Xiao:2007aa, Xiao:2012aa}).

However, such weak fields 
pose a challenge
for switching the generated excitations at the ultrafast time-scales, desirable due to short valley lifetimes~\cite{Vitale:2018aa} ($\sim 10^3-10^6$ fsec for excitons and electrons, respectively). A  
major step towards meeting this challenge has been 
made recently ~\cite{Langer:2018aa}: 
switching of the population 
between the $\mathbf{K}$ and $\mathbf{K'}$ valleys 
was achieved 
using the combination of a resonant pump pulse, 
which populated the 
desired valley, and a strong terahertz  
pulse, which moved the excited population within the Brillouin zone by 
controlling the THz field strength~\cite{Langer:2018aa}.

The feasibility of applying strong non-resonant fields  
to bulk dielectrics~\cite{Ghimire:2010aa, Schiffrin2012, Garg2016, Hohenleutner2015} and ultrathin transition methal dichalcogenide (TMD) films without material damage~\cite{Liu2016} opens major new opportunities in valleytronics. Our approach capitalizes on them. 
In contrast to previous work~\cite{Langer:2018aa}, we require neither resonant light, nor the precise tuning of
the strength of the control field.  Instead, we use far off-resonant light to modify the topological properties of the 
system by inducing Haldane-type ~\cite{Haldane1988} complex-valued second neighbour hoppings in a topologically-trivial  lattice.
This is done by using the
bicircular light field composed of counter-rotating fundamental and its second harmonic. We find that rotating the Lissajous figure drawn by the electric field vector of such pulse (Fig.~\ref{fig:bandmodification}b),  relative to the lattice (Fig.~\ref{fig:bandmodification}c),
controls the magnitude and the phase of the complex light-induced second-neighbour hoppings, thus controlling the cycle-averaged band structure (Fig.~\ref{fig:bandmodification}d) and the Berry curvature in each valley.
Exponential sensitivity of multi-photon excitation to the effective bandgap naturally leads to selective excitation in the valley where the bandgap is reduced.
Thus, valley selection is achieved by tailoring the 
symmetry of the Lissajous figure
to the 
lattice and controlling its orientation. 

Using light to control topological properties 
of solids has led to the concept of topological Floquet 
lattices 
~\cite{Oka2009,Lindner2011,Dutreix2016}. 
In this context, in addition to providing a
real-space description of the effect,
our key results are as follows.
First, we find that strong low-frequency circularly polarized 
fields show opposite valley polarization than those in the weak-field, one-photon resonant regime, as a consequence of light-induced streaking of the excited electrons. 
Second, we initialize and manipulate valley polarization on 
a few femtosecond time scale in a way that 
remains consistent for a broad range of frequencies and field intensities, and independent of the 
specifics of the material. We give two examples, hexagonal boron nitride (hBN) and MoS$_2$.
Third, using an additional linearly polarized probe pulse, we  
map the valley pseudospin onto the polarization of its harmonics, providing an all-optical measure 
of the valley asymmetry. Finally, we show numerical evidence of a topological phase transition induced by non-resonant, tailored light, occurring at specific values of intensity and wavelength of the driving field, just as predicted by our analytical 
model.

\begin{figure}
\centering
\includegraphics[width=\linewidth]{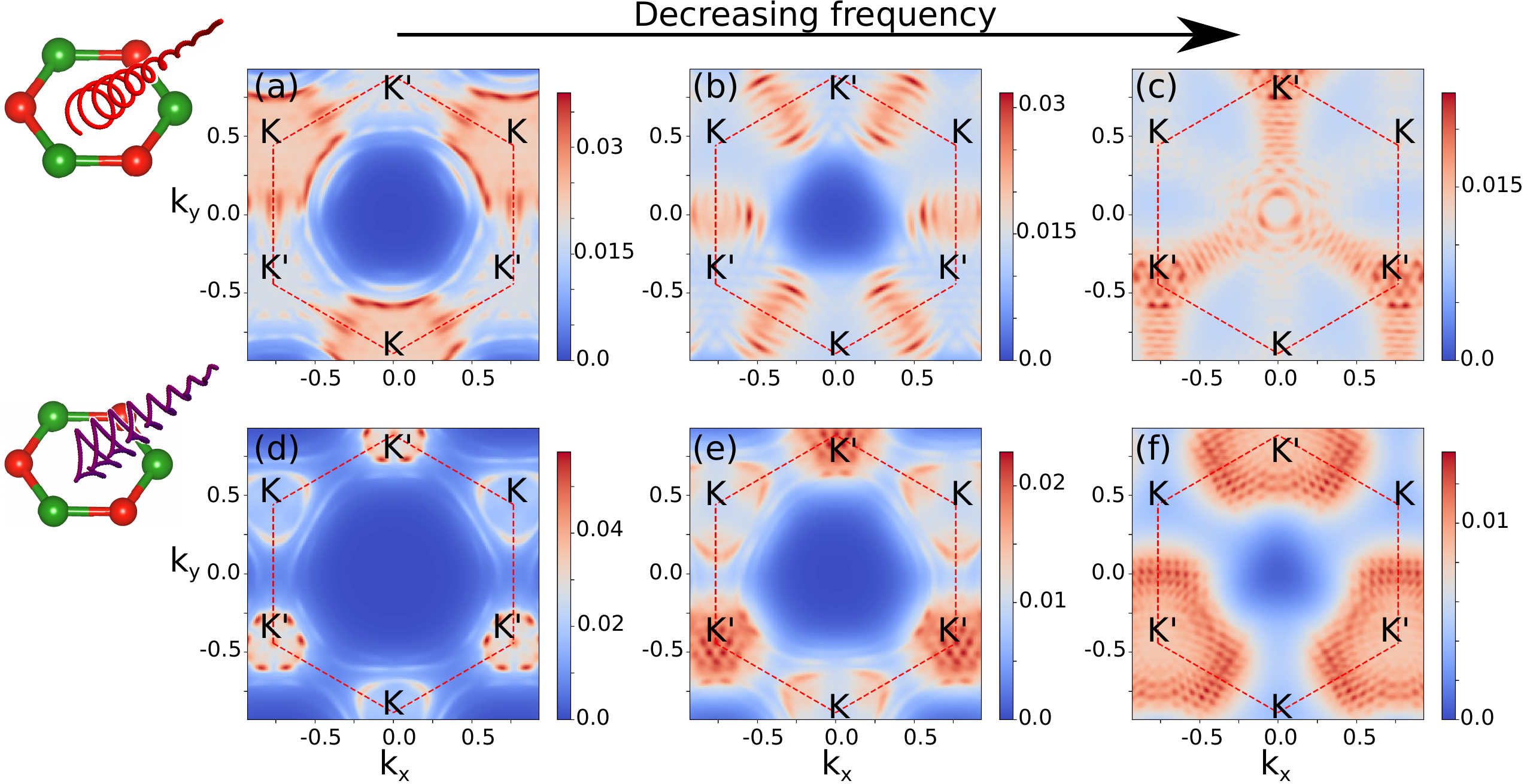}
\caption{\textbf{Selective valley excitation in strong fields}.  (a-c) Populations in the first Brillouin zone (red dashed hexagon) of the conduction band of monolayer hBN (band gap of $\Delta=5.9$~eV) after applying a strong, circularly polarized 
pulse with frequencies: (a) $\omega = 0.95$~eV, (b) $\omega = 0.68$~eV and (c) $\omega = 0.41$~eV. All other parameters (helicity, peak field strength $F_L = 0.012$~a.u. and duration $\tau = 200$~fs) remain fixed. Valley selection changes in the low-frequency regime; all panels switch $\mathbf{K}$ for $\mathbf{K'}$ when the helicity is reversed (not shown). Panels (d-f) show the same 
for a bicircular field with the same $F_L$, $\tau$ and fundamental frequency $\omega$ as in the corresponding 
panels (a-c) above. The most populated valley now remains the same; reversing the overall helicity of the fields does not switch $\mathbf{K}$ to $\mathbf{K'}$ (not shown), indicating new 
robust mechanism for valley polarization.
\label{fig:SFValley}}
\end{figure}

Consider first strong circularly polarized fields with frequencies well below the band gap energy. In Fig.~\ref{fig:SFValley}a,b,c we show the electron populations in the $p_z$ conduction band of hBN after applying a strong ($I=5$~TW/cm$^2$) circularly polarized field with three different frequencies $\omega$ and the same helicity. The same observable can be obtained, e.g., by angularly-resolved photoemission spectroscopy (ARPES). The most excited valley switches as we transition from the highest frequency (
Fig.~\ref{fig:SFValley}a) to the lowest frequency (
Fig.~\ref{fig:SFValley}c). All panels switch $\mathbf{K}$ for $\mathbf{K'}$ when the helicity of the laser is reversed (See Supplementary Note 1).

This switch in valley polarization is a consequence of streaking: the rotation of the field selects excitation at the  $\mathbf{K}$ valley, as in the one-photon case, while the large magnitude of the vector potential ($A_0 = \sqrt{I}/\omega \simeq 0.8$~a.u.) displaces the electron population towards the $\mathbf{K}'$ valley. 
%
Thus, one can 
control valley polarization using non-resonant, low-frequency fields and controlling the field helicity.
However, just like in one-photon resonant fields, the result  is material specific and depends crucially on the relation between the band gap and 
the field frequency (Fig.~\ref{fig:SFValley}a,b,c). 

A robust approach is offered by a field widely 
used for controlling strong-field processes in atoms ~\cite{Eichmann1995, Pisanty2017, Kfir2014}, which combines 
circularly polarized fundamental $\omega$ 
with its counter-rotating second harmonic:
\begin{eqnarray}
\mathbf{F}_L = \hat{\mathbf{x}} \left[-F_1 \cos(\omega t) + F_2 \cos(2\omega t + \varphi) \right] + 
\hat{\mathbf{y}} \left[F_1 \sin(\omega t) + F_2 \sin(2\omega t + \varphi) \right],
\end{eqnarray}
where $F_1$ and $F_2$ are the field strengths of the fundamental and second harmonic respectively, $\varphi$ is the sub-cycle 
phase delay between the two drivers. During one cycle, the field draws the trefoil shown in Fig~\ref{fig:bandmodification}b, which fits ideally the geometry of two triangular sub-lattices and can break the symmetry between these two identical sub -lattices in graphene  inducing charge oscillations between them ~\cite{Nag2019}. Its orientation relative to the sublattices is controlled by $\varphi$.  

Fig.~\ref{fig:SFValley}d,e,f shows the valley polarization in hBN at fixed $\varphi$, for three different frequencies. In contrast to (a-c), now the most populated valley remains robust as 
we transition from the few-photon to the deep multiphoton regime (Fig.~\ref{fig:SFValley}d,e,f). It is selected 
by the orientation of the trefoil relative to the lattice.
The feasibility of changing the trefoil orientation 
during the pulse implies the ability to switch 
the valley pseudospin on the fly.
Reversing the helicity of the two drivers ($\omega$ and $2\omega$) does not switch the valley polarization (see Supplementary Note 1), suggesting a new, robust mechanism for valley selection. 

As shown in the Methods section, energy conserving processes involving both $2\hbar \omega$ and $\hbar \omega$ photons, 
such as absorption of one $2\hbar\omega$ photon and re-emission of two $\hbar \omega$ photons, lead to
complex second-neighbour hopping $t_2$. 
Its amplitude and phase are controlled by the 
field strengths and the two-color phase $\varphi$.  
A non-zero imaginary part lifts the valley degeneracy. For moderately strong fields (see Methods), the cycle-averaged, the imaginary component of laser-induced second neighbour hopping is
\begin{equation}
\Im\{t_2\} \sim 2 J_2\left(\frac{\sqrt{3}\,a_0\,F_1}{\omega}\right) J_1\left(\frac{\sqrt{3}\,a_0\, F_2}{2\omega}\right) \cos \varphi,
\end{equation}
where $a_0$ is the lattice constant and $J_n$ is the Bessel function of the first kind of order $n$. Thus, the sub-cycle control over the field geometry, $\varphi$, controls the topological properties of the dressed system. The
associated modification of the band structure leads to 
the valley asymmetry, controlled by 
the two-color delay $\varphi$ 
in a way that is 
not material specific. 

To illustrate this, we consider monolayer hBN (see Methods for numerical details) and use a bicircular pulse with fundamental frequency $\omega = 0.41$~eV ($\lambda = 3000$~nm), duration of $\tau = 200$~fs, intensity ratio of $I(\omega)/I(2\omega) = 4$, and maximum peak intensity of $I = 5$ ~TW/cm$^2$ (lower than its predicted damage threshold~\cite{Tancogne2018}). We change the two-color phase $\varphi$ to control the orientation of the field with respect to the lattice.
The valley population asymmetry is calculated as $\mathcal{A} = \pm 2(f_{n,K}-f_{n,K'})/(f_{n,K}+f_{n,K'})$, where $f_{n,K}$ ($f_{n,K'}$) is obtained by integrating the electron population inside the black dashed circles encircling $\mathbf{K}$ ($\mathbf{K'}$) in (a-c,e-g), and the $+$ ($-$) sign is used for hBN (MoS$_2$), due to their opposite Berry curvatures.  

\begin{figure}
\centering
\includegraphics[width=\linewidth]{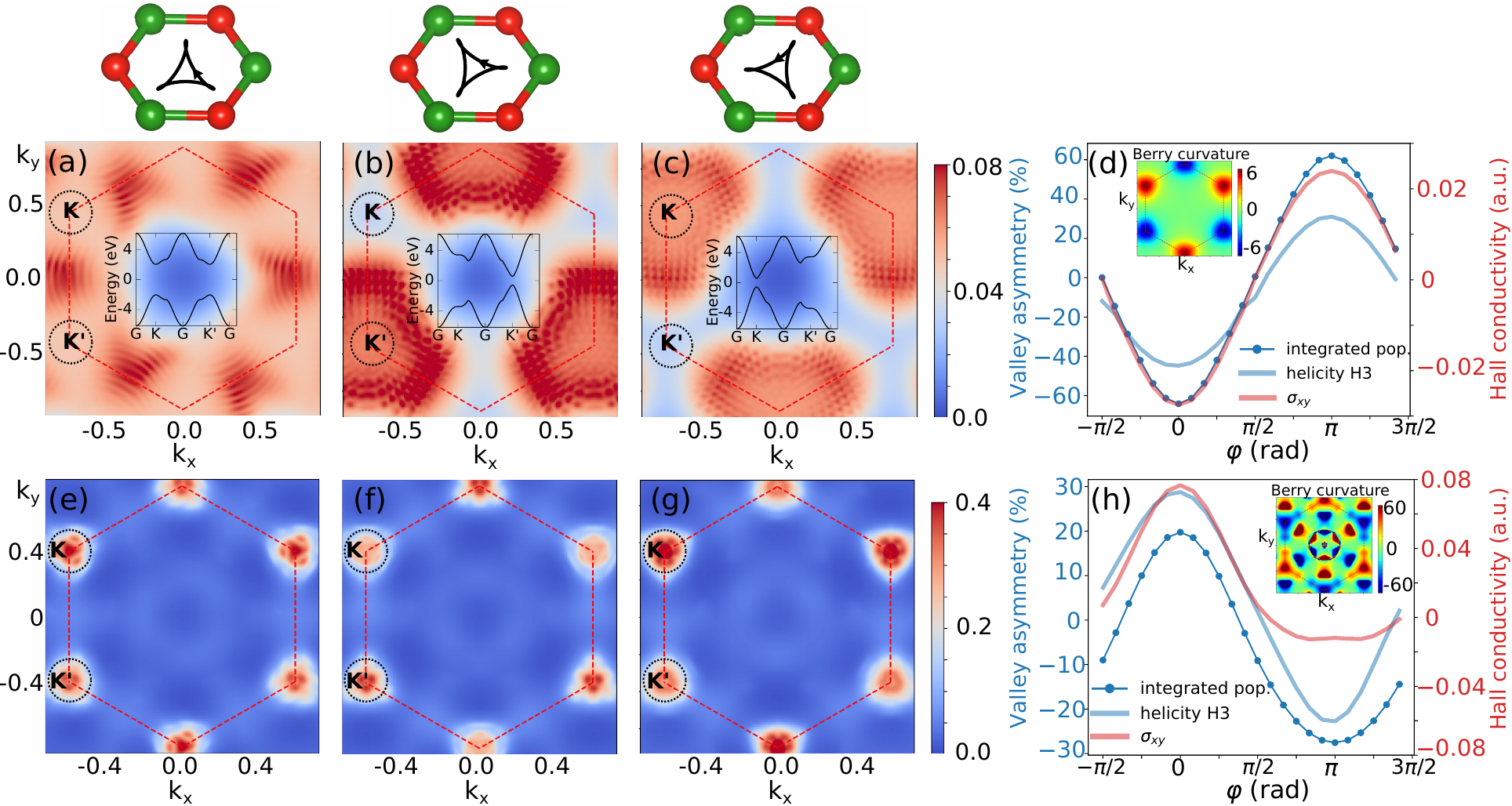}
\caption{\textbf{Strong-field manipulation and optical reading of valley polarization}.  Electron populations in the lowest conduction bands of hBN (a-c) and  spin-integrated MoS$_2$ (e-g), after applying a bicircular field with $\lambda=3\mu$m:
(a,e) $\varphi = -\pi/2$, (b,f) $\varphi = 0$, and (c,g) $\varphi = \pi$, sketches of field orientations 
relative to the hexagonal cell are also shown. 
The insets in panels (a)-(c) show the cycle-averaged band structures for hBN (see Methods) vs $\varphi$.
(d,h) Valley population asymmetry (dotted solid blue curve), helicity of the third harmonic of the probe (faint blue curve) and anomalous Hall conductivity (faint red curve), as a function of $\varphi$ 
in (d) hBN and (h) MoS$_2$. 
The probe is carried at $3\omega$ for hBN and $\omega$ for MoS$_2$; $\omega$ is the fundamental frequency of the bicircular pulse. 
\label{fig:manipulation}}
\end{figure}

When $\varphi = -\pi/2$, both sub-lattices are addressed 
equally, no valley asymmetry is present in 
the cycle-averaged band structures (Fig.~\ref{fig:manipulation}a). We find that $\mathbf{K}$ and $\mathbf{K'}$ valleys of the conduction band are nearly equally excited. In contrast, when $\varphi = 0$, the cycle-averaged band structure shows a strong valley asymmetry (Fig.~\ref{fig:manipulation}b). This reflects in the valley populations, which show a 60\% contrast (Fig.~\ref{fig:manipulation}d). 
The situation is reversed for $\varphi = \pi$, the populations switch to the opposite valley (Fig.~\ref{fig:manipulation}c). The sense of rotation of the pulse also contributes to the valley asymmetry due to the orbital propensity rule, but its effect is relatively weak. It manifests in small valley polarization for $\varphi=\pm\pi/2$, and in that $\varphi=0$ and $\varphi=\pi$ are not exact opposites (compare Figs.~\ref{fig:manipulation}b,c).

To confirm that the mechanism is general, we performed calculations in MoS$_2$ for the same laser frequencies, which allows us to maintain the same timescale of valley control. Due to the smaller band gap of MoS$_2$, the intensity peak of the total field was kept at $I = 0.15$~TW/cm$^2$, well below its damage threshold~\cite{Liu2016}. 
Fig.~\ref{fig:manipulation}(e)-(g) shows valley excitation after the pulse in the spin-integrated lowest conduction band of MoS$_2$ (see Supplementary Note 2 for spin-resolved results), illustrating similar control as for hBN, but with lower values of the valley polarization. Higher values of valley polarization can be obtained by increasing the field strength, while total populations can be controlled with the pulse duration.

We now turn to optical reading of the valley pseudospin. Since the Berry curvature is opposite at the $\mathbf{K}$ and $\mathbf{K'}$ valleys, when an in-plane electric field is applied, the carriers generate a current perpendicular to the electric field (anomalous current), with opposite direction at each valley. For equal valley populations, the anomalous current will thus cancel, leading to a zero anomalous Hall conductivity (AHC),
\begin{equation}\label{eq:AHC}
\sigma_{xy} = -\frac{e}{\hbar}\,\sum_n \int_{\text{BZ}} \frac{d\mathbf{k}}{(2\pi)^3} f_n (\mathbf{k}) \Omega_{n,z} (\mathbf{k}), 
\end{equation}
where $f_n$ and $\Omega_n$ are the population and Berry curvature of the $n$-th band. 
If the valley populations are not equal, the perpendicular currents originating from $\mathbf{K}$ and $\mathbf{K}'$ do not compensate each other, leading to a non-zero AHC~\cite{Xiao:2012aa}.  This so-called valley Hall effect has been demonstrated for MoS$_2$ monolayers by measuring the direction of the transverse Hall voltage~\cite{Mak:2014aa}. 

Alternatively, the direction of the anomalous current 
can be retrieved all-optically from the helicity of the harmonics of a linearly polarized probe ~\cite{Silva2019Topo}. To this end,
we apply a probe field with frequency $\omega$ or 
$3\omega$, linearly polarized along the $\mathbf{G}$-$\mathbf{M}$ direction of the lattice, 
and monitor the harmonic response at a 3N multiple of the probe frequency. This guarantees a background-free measurement without interference from harmonics generated by the bi-circular field, since the latter does not generate $3N\omega$ harmonics due to symmetry. The probe pulse arrives after the end of the bicircular pump. 

The component of the optical response 
polarized orthogonal to the probe polarization 
undergoes a phase jump of $\pi$ as the population switches from one valley to another (see Methods), leading to a rotation in the helicity of the probe response,
Fig.~\ref{fig:manipulation}(d,h).
The helicity is calculated as $h = 2(I_{\lcirclearrowright} - I_{\rcirclearrowleft})/(I_{\lcirclearrowright} + I_{\rcirclearrowleft})$, where  $I_{\lcirclearrowright}$ ($I_{\rcirclearrowleft}$) is the component of the harmonic intensity rotating  clockwise (anticlockwise).

When the dynamics occur mainly in one conduction band, as in monolayer hBN, the helicity of the optical response reads the AHC and, consequently, the valley polarization (Fig.~\ref{fig:manipulation}d). In the case of MoS$_2$ (Fig.~\ref{fig:manipulation}h), the helicity follows the valley polarization qualitatively, but deviates from 
the AHC due to the influence of higher bands. 

\begin{figure}
\centering
\includegraphics[width=\linewidth]{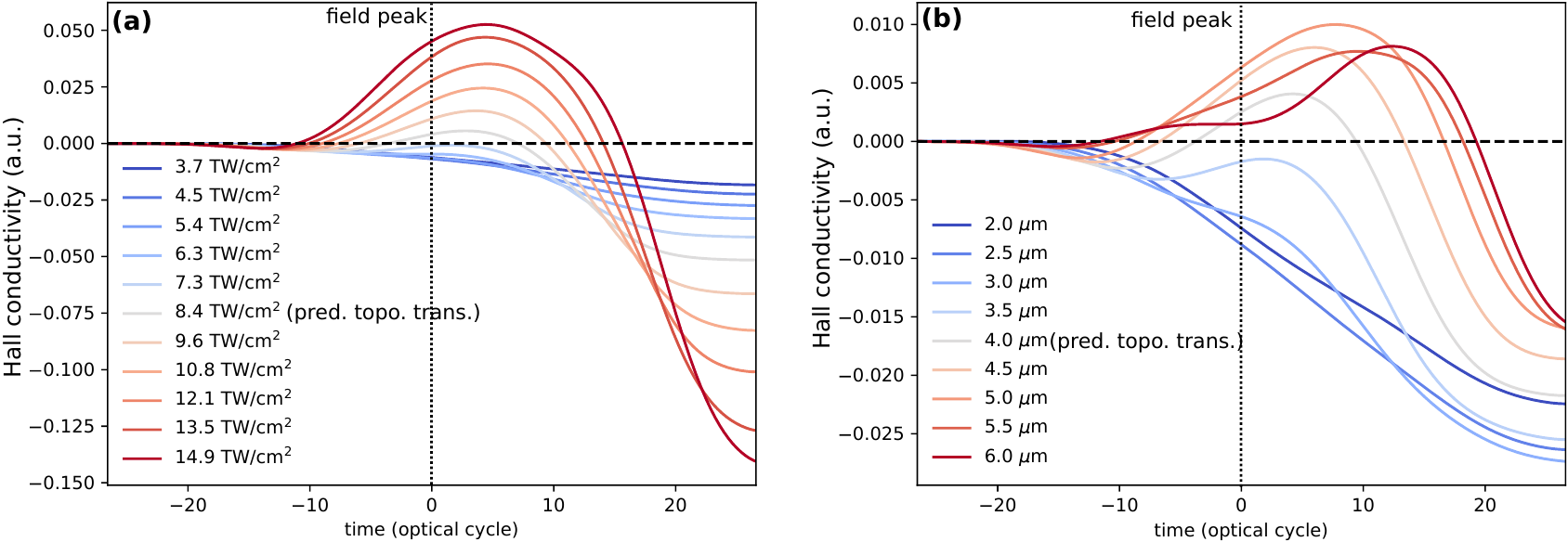}
\caption{\textbf{Light-induced topological phase transition}. (a,b) Gated time-dependent anomalous Hall conductivity (AHC) in hBN, for bi-circular field with (a) different total electric field amplitude 
for fixed carrier $\lambda=3\mu$m, and (b) different $\lambda$ 
for fixed total intensity $I_{\rm tot}=5$~TW/cm$^2$. The pulse duration is 20 cycles (FWHM),
$I({\omega})/I({2\omega}) = 4$,  two-color delay $\varphi=0$. 
The grey lines at the total intensity
$I_{\rm tot}=8.4$~TW/cm$^2$ 
(a) and $\lambda = 4\mu$m (b) correspond to the 
analytical prediction of the topological phase 
transition. It 
coincides with the change of sign of the gated AHC close to the peak of the field (warm color lines). 
\label{fig:topology}}
\end{figure}

Our approximate analytical model for an effective 
$t_2$ Eq.(2) predicts a topological phase transition. 
To explore this prediction, we monitor 
the time-dependent anomalous Hall conductivity (AHC) 
in hBN during the laser pulse.   
Fig.~\ref{fig:topology} shows the gated time-dependent AHC defined as
\begin{equation}
    \bar{\sigma}_{xy} (t)= \int_{-\infty}^{\infty} \sigma_{xy}(t')\,G(t'-t) dt',
\end{equation}
where $\sigma_{xy}(t')$ is the instantaneous AHC, defined by Eq.~\ref{eq:AHC} using
instantaneous electron populations $f_n (\mathbf{k},t)$.
The instantaneous AHC is averaged over several laser cycles using the gate function $G(t'-t) = e^{-4 \ln 2 (t'-t)^2/T^2}$, $T = 4\times 2\pi/\omega$, to produce $\bar{\sigma}_{xy} (t)$. 

Note that $\sigma_{xy} (t)$ uses field-free Berry curvature 
$\Omega (\mathbf{k})$ and field-free bands, for which $f_n (\mathbf{k},t)$ are computed. 
Strikingly, we find that $\bar{\sigma}_{xy} (t)$ changes sign when 
the laser field strength and wavelength reach the parameter regime 
for the 
topological phase transition predicted by the model, both as a function of intensity and frequency (Fig.~\ref{fig:topology}a,b). 
The sign change in $\sigma_{xy} (t)$ manifests the key feature of the topological phase transition, that is, the band gap closing and the
associated change in the cycle-averaged anomalous Hall conductivity of the dressed system. 

Thus, intense low-frequency fields offer unusual and robust routes for controlling electronic response in 2D materials at PHz rates.  
Fields with polarization states 
tailored to the geometry of the lattice 
provide the most opportunities. Controlling 
orientation of the field polarization relative 
to the lattice controls the topological properties 
of the trivial dielectric by inducing and 
controlling complex Haldane-type second-neighbor 
couplings. This enables robust valley selection 
for multi-photon excitation.
The new rules for valley selective excitation
could also enable spin selectivity in materials with strong spin-orbit interaction, exploiting exponential sensitivity
of strong-field excitation to the bandgap
in the way similar to spin-polarization in strong-field
ionization of atoms \cite{barth2013spin,hartung2016electron}.
Our work thus lays the grounds for a new regime of valleytronics, spintronics, and light-induced topology.

\section*{References}

\bibliographystyle{naturemag}
\bibliography{biblio}

\end{document}